\documentclass[a4paper]{jpconf}
\usepackage{graphicx}

\usepackage{mathtools}
\usepackage{amsmath}
\usepackage{graphicx,booktabs,array}
\usepackage{mwe} 
\usepackage{subcaption}
\usepackage{listings}
\usepackage{physics}
\usepackage{appendix} 

\usepackage{xcolor}

\definecolor{codegreen}{rgb}{0,0.6,0}
\definecolor{codegray}{rgb}{0.5,0.5,0.5}
\definecolor{codepurple}{rgb}{0.58,0,0.82}
\definecolor{backcolour}{rgb}{0.95,0.95,0.92}

\lstdefinestyle{mystyle}{
  backgroundcolor=\color{backcolour}, commentstyle=\color{codegreen},
  keywordstyle=\color{magenta},
  numberstyle=\tiny\color{codegray},
  stringstyle=\color{codepurple},
  basicstyle=\ttfamily\footnotesize,
  breakatwhitespace=false,         
  breaklines=true,                 
  captionpos=b,                    
  keepspaces=true,                 
  numbers=left,                    
  numbersep=5pt,                  
  showspaces=false,                
  showstringspaces=false,
  showtabs=false,                  
  tabsize=2
}

\begin{document}

\title[]{Simulating molecules using the  VQE algorithm on Qiskit }


\author{Alan Anaya and Francisco Delgado}

\address{{Tecnologico de Monterrey, School of Engineering and Sciences, 
M\'exico.}}
\ead{fdelgado@tec.mx}

\section*{Introduction}
Feynman's ideas to employ quantum systems  for simulating other quantum systems \cite{Feynman} gave rise to quantum simulation \cite{Lloyd1}. 
While current quantum computers are still prone to decoherence and rely on error correction \cite{Ladd}, the development of  hybrid algorithms that employ both quantum and classical computation allow to perform quantum simulation  \cite{Jarrod}. 

Among these algorithms, the Variational Quantum Eigensolver (VQE) algorithm has permitted to explore the electronic structure of simple atoms and molecules by exploiting the  Rayleigh–Ritz variational principle \cite{Peruzzo}. This has enabled theoretical chemists to gain insight about reaction
rates, find  chemically stable structures, determine spectroscopic properties of molecules, among others.

The VQE employs a parameterized circuit that prepares a quantum state that aims to approach the ground-state of the molecular Hamiltonian. The expectation value of this Hamiltonian, calculated by a procedure known as Hamiltonian averaging  \cite{McClean}, provides a cost function to minimize by optimizing the set of parameters. This optimization procedure is performed using a classical computer. These steps are performed iteratively until convergence is found. Among some of the main challenges of VQE is it's dependence on the variational form to find the global minimum for the energy expectation value. Both hardware efficient variational forms \cite{Kandala1}, that employ a limited set of gates, and chemically inspired  \cite{Romero} that adapt classical computational chemistry methods and take into account the chemical details, have been proposed. 

In this supplementary material, we introduce the technical application of VQE in quantum chemistry by calculating the ground-state single point energy of the $H_2$ molecule. Furthermore, we explore the Energy potential surface over a range of distances and discuss the effect of the variables on the accuracy of the results. Together, we provide the implementation of the VQE algorithm for finding the ground-state energy of the H\textsubscript{2} molecule on Qiskit \cite{qiskit} library for python. 

\section{Dictionary of terms}\label{appendix}
In this section we define the dictionary of terms to be employed on the python code. This allows reducing the notation and clarifying the code. The dictionary of terms include the quantum instance to be used for executing the quantum circuit, the required mapping for measuring the molecular Hamiltonian's expectation value and the classical optimizers to be used for the parameter minimization.

\begin{lstlisting}[language=Python]
# Simulators
SV = "statevector"
QA = "qasm"

# Mappings
PA = "parity"
JW = "jordan_wigner"
BK = "bravyi_kitaev"

#Optimizers
CO = "COBYLA"
BF = "L_BFGS_B"
SL = "SLSQP"
SP = "SPSA"

# Quantum Instances for running the quantum circuit
quantum_instances = {
    SV: QuantumInstance(backend=BasicAer.get_backend('statevector_simulator')),
    QA: QuantumInstance(backend=Aer.get_backend('qasm_simulator'))}
    
# Mappings for transforming the molecular Hamiltonian into a set of Pauli products acting on qubits
map_types = {
    PA: "parity",
    JW: "jordan_wigner",
    BK: "bravyi_kitaev"}
    
# Classical Optimizers for parameter minimization
MAX_ITER = 500
optimizers = {
    CO: COBYLA(maxiter=MAX_ITER),
    BF: L_BFGS_B(maxiter=MAX_ITER),
    SL: SLSQP(maxiter=MAX_ITER),
    SP: SPSA(max_trials=MAX_ITER,save_steps=100)}
\end{lstlisting}

\section{Functions}\label{appendixB}
In the current section, we provide a series of functions which will allow to implement the VQE algorithm. We separated different stages of the implementation in the following form: \vspace{0.5cm}

\noindent 1) A function generating the qubit operators of the molecular Hamiltonian with the relevant information of the system (number of particles, number of spin orbitals and energy shift) given as inputs the inter-atomic distance between hydrogen atoms, the basis set  to perform the one and two body integrals and the map to be implemented. If you desire to get the qubit operator details set verbose to true. The two qubit reduction (tqr) is set to false and can be used to reduce the number of qubits by exploiting the symmetries of the Hamiltonian \cite{parity,Aspuru}.

\begin{lstlisting}[language=Python]
def get_qubit_op_H2(dist, basis, map_type, verbose=False,tqr=False):
    driver = PySCFDriver(atom="H .0 .0 .0; H .0 .0 " + str(dist), unit=UnitsType.ANGSTROM, charge=0, spin=0, basis=basis)
    molecule = driver.run()
    #Calculate nuclear repulsion energy
    repulsion_energy = molecule.nuclear_repulsion_energy 
    num_particles = molecule.num_alpha + molecule.num_beta
    num_spin_orbitals = molecule.num_orbitals * 2
    #Calculate one and two body integrals
    h1=molecule.one_body_integrals
    h2=molecule.two_body_integrals
    ferOp = FermionicOperator(h1=h1, h2=h2) 
    #Perform the Mapping from Fermionic operators to Qubit operators
    qubitOp = ferOp.mapping(map_type=map_type, threshold=0.00000001)
    shift =  repulsion_energy
    if verbose:
        print(h1)
        print(h2)
        print(qubitOp)
        print(qubitOp.print_details())
    if tqr:
        qubitOp = Z2Symmetries.two_qubit_reduction(qubitOp, num_particles)
    if verbose:
        print(qubitOp)
        print(qubitOp.print_details())
    return qubitOp, num_particles, num_spin_orbitals, shift
\end{lstlisting}

\vspace{0.5cm}

\noindent 2) A function that constructs an ansatz consisting of a variational circuit and an initial reference state. This ansatz will be used to prepare a parameterized quantum state. As inputs, the function requires the outputs obtained from the previous stage (qubit operators, number of orbitals and number of particles). 

\begin{lstlisting}[language=Python]
def build_ansatz(qubitOp, num_orbitals, num_particles, map_type="parity", initial_state="HartreeFock", var_form="UCCSD", depth=1,tqr=False):
    # Specify your initial state
    initial_states = {
        'Zero': Zero(qubitOp.num_qubits),
        'HartreeFock': HartreeFock(num_orbitals, num_particles, map_type)}
    # Select a state preparation ansatz
    # Equivalently, choose a parameterization for the trial wave function.
    var_forms = {
        'UCCSD': UCCSD(num_orbitals=num_orbitals, num_particles=num_particles,initial_state=initial_states[initial_state], qubit_mapping=map_type, reps=depth,two_qubit_reduction=tqr),
        'RealAmplitudes': RealAmplitudes(qubitOp.num_qubits, reps=depth,initial_state=initial_states[initial_state]), 
        'EfficientSU2': EfficientSU2(qubitOp.num_qubits, reps=depth,initial_state=initial_states[initial_state]), 
        'TwoLocal': TwoLocal(qubitOp.num_qubits,['ry','rz'], 'cz', 'full', reps=depth,initial_state=initial_states[initial_state]), 
        'ExcitationPreserving': ExcitationPreserving(qubitOp.num_qubits,mode='iswap', entanglement='full',reps=depth,initial_state=initial_states[initial_state])}
    return var_forms[var_form]
\end{lstlisting}

\vspace{0.5cm}

\noindent 3) A function that calculates the exact eigenvalues of the Hamiltonian calculated with NumPy. As output, it returns the lowest eigenvalue corresponding to the ground-state energy.

\begin{lstlisting}[language=Python]
def exact_solver(qubitOp, verbose=False):
    ee = NumPyEigensolver(qubitOp)
    result = ee.run()
    ref = result['eigenvalues'].real[0]
    if verbose:
        print('Reference value: {}'.format(ref))
    return ref
\end{lstlisting}

\vspace{0.5cm}

Finally, we  also provide a function that generates a random set of  parameters to be set as initial points for the variational circuit.

\begin{lstlisting}[language=Python ]
def random_initial_point(ansatz, interval=[0,1]):
    initial_point = []
    if len(interval)>1:
        for i in range(ansatz._num_parameters): # For UCCSD
            initial_point.append(random(interval[0], interval[1]))
    else:
        for i in range(ansatz._num_parameters):
            initial_point.append(random(0, interval[0]))
    return initial_point
\end{lstlisting}

\noindent and another function that saves the convergence information about the VQE execution.

\begin{lstlisting}[language=Python]
def callback(nfev, parameters, energy, stddev):
    intermediate_info['nfev'].append(nfev)
    intermediate_info['parameters'].append(parameters)
    intermediate_info['energy'].append(energy)
    intermediate_info['stddev'].append(stddev)
\end{lstlisting}

\section{VQE Implementation on Qiskit for a single point calculation}\label{appendixC}
In this section, we provide the implementation of the VQE for a single point calculation at the inter-atomic equilibrium distance of 0.74 Å. First we start by importing the required python libraries.
\begin{lstlisting}[language=Python]
# Import from Qiskit
from qiskit import BasicAer, Aer, IBMQ
from qiskit.circuit.library import RealAmplitudes, EfficientSU2, TwoLocal, ExcitationPreserving
# Import from Qiskit Aqua
from qiskit.aqua import QuantumInstance
from qiskit.aqua.algorithms import VQE, NumPyEigensolver
from qiskit.aqua.components.initial_states import Zero
from qiskit.aqua.components.optimizers import COBYLA, L_BFGS_B, SLSQP, SPSA
from qiskit.aqua.operators import Z2Symmetries
# Import from Qiskit Chemistry
from qiskit.chemistry import FermionicOperator
from qiskit.chemistry.drivers import PySCFDriver, UnitsType
from qiskit.chemistry.components.variational_forms import UCCSD
from qiskit.chemistry.components.initial_states import HartreeFock

# Import from other python libraries
import numpy as np
import matplotlib.pyplot as plt
import pylab as py
\end{lstlisting}

Then, we define the set of variables required for executing the algorithm. The basis set chosen for calculating the one and two body integrals  was a minimal basis set of 3 Gaussian orbitals fitted to a Slater Orbital, denoted as STO-3G. We choose to map the Molecular Hamiltonian to Pauli operators by using the Parity mapping, this allow us to do a two-qubit reduction process. The Hartree-Fock state was chosen as initial reference state. The variational form and  the classical optimizer can be changed to wish in order to observe the difference between methods. For this particular case, we chose to employ the Unitary Couple Cluster (UCCSD) as our variational form and the Sequential Least Squares Programming (SLSQP) optimizer. For the calculations, the variational quantum circuit is executed on a state-vector simulator.

\begin{lstlisting}[language=Python]
# 1. Set the Configuration needed to run VQE on Qiskit

# Interatomic distance.
dist = 0.74 
# Basis set for calculating the one and two body integrals.
basis='sto3g' 
# Mapping type.
map_type = map_types[PA]  
# Initial reference state to be evolved on the VQE circuit.
initial_state='HartreeFock'
# Variational form to be used.
var_form='UCCSD' 
# Quantum simulator
quantum_instance = quantum_instances[SV] 
# Classical optimizer.
optimizer=optimizers[SL] 
# Two qubit reduction set to True
tqr=True 
\end{lstlisting}

In the next step, we obtain the qubit operators, the number of particles, number of orbitals and the energy shift from the nuclear repulsion energy. 

\begin{lstlisting}[language=Python]
# 2. Get the Hamiltonian in terms of Pauli matrices and the variables to build the ansatz.

qubitOp, num_particles, num_orbitals, shift = get_qubit_op_H2(dist, basis, map_type, verbose=False,tqr=tqr)
\end{lstlisting}

Then, we construct the variational circuit to be implemented. A random initial set of points can be given to the VQE circuit. Moreover, callback of the intermediate information can be done in order to get the convergence of the parameters and the energy during the simulation.

\begin{lstlisting}[language=Python]
# 3. Bulid the ansatz for the given variational form and initial reference state.
ansatz= build_ansatz(qubitOp, num_orbitals, num_particles, map_type=map_type,  initial_state=initial_state, var_form=var_form,tqr=tqr)

# 4. Set an initial random set of parameters for the quantum circuit
initial_point = np.random.random(ansatz.num_parameters)


# 5. Save the intermediate information (number of evaluations, parameters, energies and standard deviation) of the VQE Algorithm.

intermediate_info = {
    'nfev': [],
    'parameters': [],
    'energy': [],
    'stddev': []
}
\end{lstlisting}

Finally, the VQE algorithm is executed and compared to the exact results. For setting random initial points to parameters and callback the intermediate information, write down the last two arguments.
\begin{lstlisting}[language=Python]
# 6. Run VQE on a quantum instance.

vqe = VQE(qubitOp, ansatz, optimizer,initial_point=initial_point,callback=callback)
result= vqe.run(quantum_instance)

# 7. Extract the VQE energy calculation.

vqe_result = np.real(vqe.run(quantum_instance)['eigenvalue'] + shift)

# 8. Compare the  VQE energy with the exact result.

ref=exact_solver(qubitOp, verbose=False)
print("Exact Result:", ref + shift)
print("VQE Result:", vqe_result)
\end{lstlisting}

Results of the the energy convergence and the parameter optimization can be plotted as a function of the number of iterations. In figure \ref{fig3},
 we show an example of these results for the single point calculation of $H_2$ molecule as described above-. Convergence will vary depending on the initial set of parameters.
{\centering
\begin{figure}[ht]
\begin{tabular}{cc}
\begin{subfigure}{0.48\textwidth}\centering\includegraphics[width=1\columnwidth]{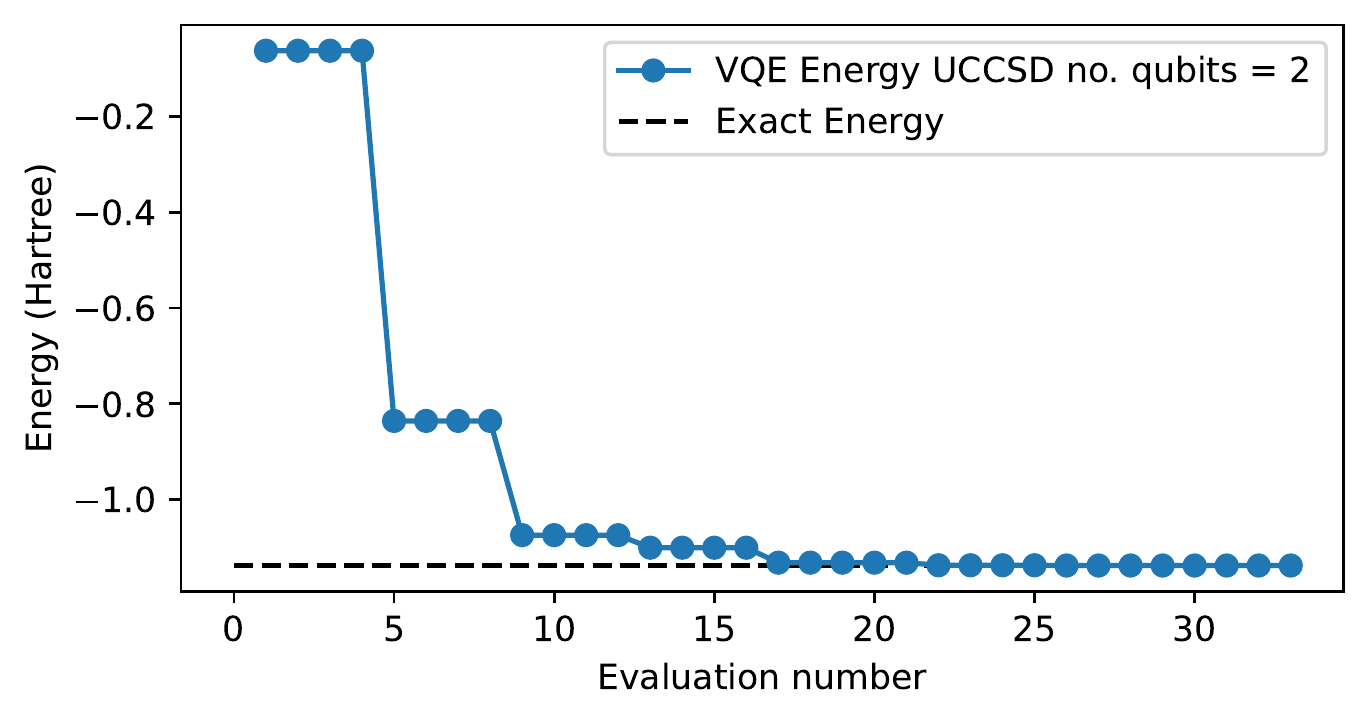}\caption{}\label{fig:taba}\end{subfigure}&
\begin{subfigure}{0.48\textwidth}\centering\includegraphics[width=1\columnwidth]{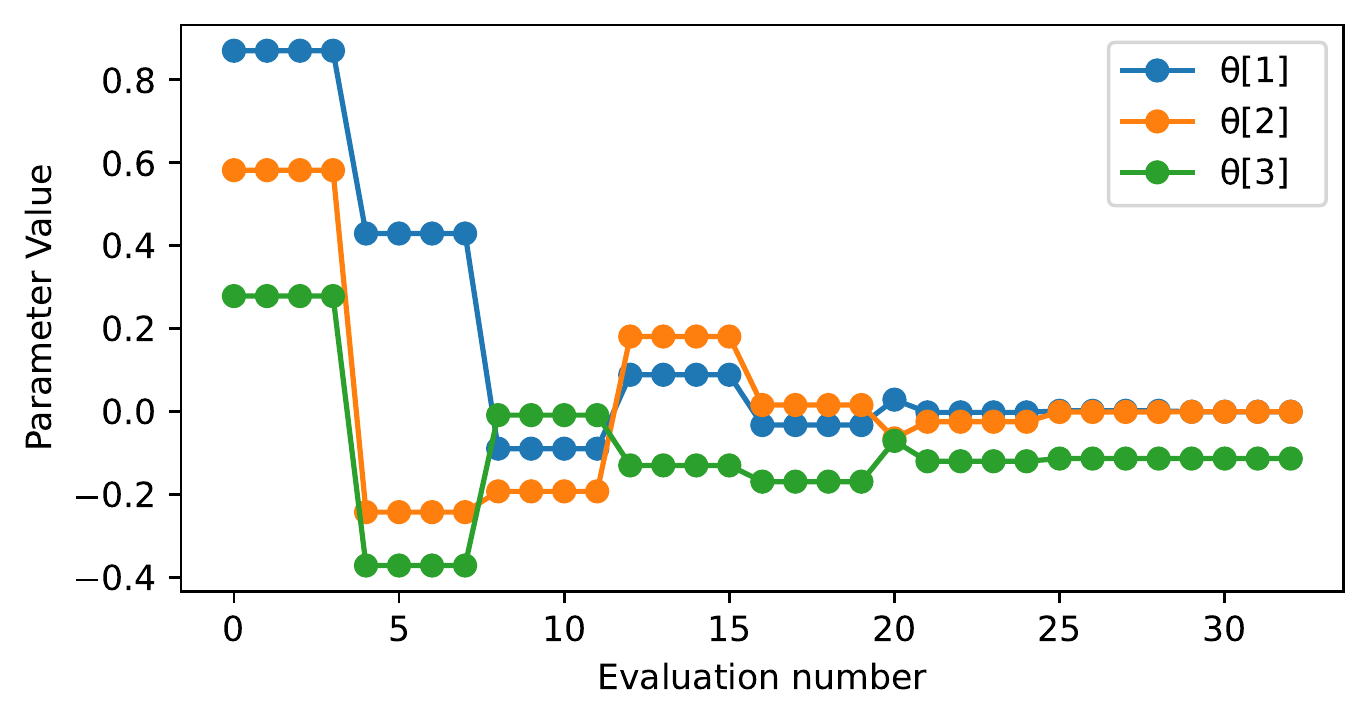}\caption{}\label{fig:tabb}
\end{subfigure}
\end{tabular}
\caption{(a) Energy convergence of the VQE implemented using the UCCSD variational form. (b) Parameter optimization of the VQE circuit.} \label{fig3}
\end{figure}
}


\section*{References}

\begin{thebibliography}{9}

\bibitem{Feynman}
Feynman R 1982 {\it Int. J. Theor. Phys.} {\bf 21} 467

\bibitem{Lloyd1}
Lloyd S 1996 {\it Science} {\bf 273} 1073

\bibitem{Ladd}
Ladd, T, Jelezko F, Laflamme R et al 2010 {\it Nat.} {\bf 464} 45

\bibitem{Jarrod}
Jarrod R McClean et al 2016 {\it New J. Phys.} {\bf 18} 023023
 
\bibitem{Peruzzo}
Peruzzo A, McClean J and Shadbolt P et al 2014  {\it Nat. Commun.} {\bf 5} 4213

\bibitem{McClean}
McClean J  et al 2016 {\it New J. Phys} {\bf 18} 023023

 \bibitem{Kandala1}
 Kandala A, Mezzacapo A, Temme K et al 2017 {\it Nat.} {\bf 549} 242
 

\bibitem{Romero}
Romero J et al 2018 {\it Quant. Sci. Technol.} {\bf 4} 014008


\bibitem{qiskit}
Gadi Aleksandrowicz et al 2019 Qiskit: An Open-source Framework for Quantum Computing (0.7.2) Zenodo https://doi.org/10.5281/zenodo.2562111

\bibitem{parity}
Bravyi S, Gambetta J,  Mezzacapo A,  and Temme K 2017 {\it arXiv: Quantum Phys.}


\bibitem{Aspuru}
McArdle S,  Endo S, Aspuru-Guzik A,  Benjamin S,  Yuan X 2020  {\it Rev. Mod. Phys.} {\bf 92} 15003







\end{thebibliography}
\end{document}